\documentclass[%
reprint,
amsmath,amssymb,
aps,
pra,
]{revtex4-2}

\usepackage{graphicx}
\usepackage{dcolumn}
\usepackage{bm}
\usepackage{mathtools}
\usepackage{atbegshi}
\usepackage{xcolor}
\usepackage{amsmath}
\usepackage{hyperref}
\hypersetup{
    colorlinks=true,
    linkcolor=blue,
    filecolor=blue,      
    urlcolor=blue,
    citecolor=blue
}

\DeclareMathOperator{\sinc}{sinc}

\begin{document}

\preprint{APS/123-QED}

\title{Tailoring indistinguishability of photons using longitudinal spatial coherence}

\author{Preeti Sharma$^{1}$}

\author{Gaytri Arya$^{1}$}

\author{Bhaskar Kanseri $^{1,2}$}
\email{bkanseri@physics.iitd.ac.in}
\affiliation{%
 $^{1}$Department of Physics, Indian Institute of Technology Delhi, Hauz Khas, New Delhi 110016, India}%
\affiliation{
  $^{2}$Optics and Photonics Centre, Indian Institute of Technology Delhi, Hauz Khas, New Delhi 110016, India
}%


\date{\today}

\begin{abstract}
Methods to generate photons with tailored indistinguishability are central to developing photonic quantum technologies and making fundamental tests of quantum physics.  
This study introduces a novel method for manipulating effective longitudinal spatial coherence (LSC) of biphotons, controlling their indistinguishability in a significant manner. The experimental results show that, instead of tailoring the frequency spectrum of the interfering photons, changing their LSC also leads to controlling the width of Hong-Ou-Mandel 
dip as validated by the theoretical calculations. This powerful approach not only modifies the conventional wisdom claiming only frequency width responsible for indistinguishability control of photons but also positions the LSC as a promising tool for fine-tuning the longitudinal coherence of photons, thereby expanding their potential use in quantum science and technologies.
\end{abstract}

\maketitle


In the realm of quantum technologies, achieving indistinguishable photons and atoms is crucial for implementing advanced quantum algorithms and protocols for quantum communication \cite{kawachi2005computational,hochrainer2022quantum}. This pursuit drives extensive research into effective methods to harness these quantum resources, promising breakthroughs in computing and information processing. Quantum particles, including bosons and fermions, play pivotal roles in various quantum technologies such as quantum computing and metrology \cite{barbarino2019engineering,wilson2020observation}. Surface plasmon polaritons (SPPs), exhibiting bosonic characteristics, are fundamental in ultracompact quantum circuitry \cite{di2014observation}. Spontaneous parametric down-conversion (SPDC) is a critical technique for producing identical photons in multiple degrees of freedom (DoF) like polarization, energy, and orbital angular momentum \cite{hong2023hong,thomas2010measurement,bouchard2020two}. Researchers are exploring new DoFs, such as radial transverse modes and twisted phase \cite{karimi2014exploring,hutter2021partially}, to expand the dimensionality of indistinguishable photons.

The Hong-Ou-Mandel (HOM) interferometer, a fundamental tool in these studies, has been instrumental in investigating the indistinguishability of photons and quantum particles \cite{lim2005generalized,kashi2024}. In the HOM setup, when two incoming photons at the two ports of a beam splitter are indistinguishable in all degrees of freedom, a characteristic HOM dip is observed in the coincidence events, signifying overlap of their wavepackets \cite{kim2020hong,siltanen2021engineering,devaux2020imaging}. The realization of the HOM effect with entangled photons has played a vital role in quantum information technologies \cite{kok2007linear,yuan2010entangled} and quantum computation \cite{wang201818}.  
The precise time measurements with HOM setup make it a potential candidate for accurate clock synchronization \cite{quan2016demonstration}. Given these compelling applications, the manipulation of the HOM dip remains a focal point of interest for assessing photon indistinguishability and delving into fascinating quantum phenomena \cite{nasr2008ultrabroadband,gera2024hong}. In previous studies, the manipulation of the HOM dip in photons generated using SPDC has been achieved by adjusting the temporal and spatial correlations in either the pump beam or the biphotons themselves \cite{drago2024hong}. 
The spectral properties of SPDC photons have played a crucial role to alter the coherence width of photons in the longitudinal direction. Traditionally, the full-width half maximum (FWHM) of the HOM dip gives the information of spectral width in generated photons. However, the coherence in the longitudinal direction is a combined effect of the longitudinal spatial coherence (LSC) and temporal coherence \cite{ryabukho2009longitudinal,rosen1995longitudinal,ryabukho2004influence,arya2021modulations}, with the effective coherence length ($L_c$) in the longitudinal direction determined using the relation $1/L_c = 1/l_c + 1/\rho_{\parallel}$, where $\rho_{\parallel}$ and $l_c$ are LSC length and temporal coherence length, respectively. Greater angular width results in smaller LSC length ($\rho_{\parallel}=2 \lambda/ \theta^2$) and higher resolution \cite{ryabukho2009longitudinal}, which has led to LSC becoming a popular technique in full-field optical coherence tomography (FF-OCT) \cite{abdulhalim2006competence,abdulhalim2013low} and other imaging schemes \cite{liu2008spatial,takeda2005coherence,Deng_2024}.

The conventional wisdom about the relation of the dip width (FWHM) of the HOM interference pattern to the spectral properties of the interfering photons is incomplete. In this letter, we illustrate the significant impact of LSC on the modification of HOM dip, providing an alternate way to control and understand quantum interference. The objective of this study is thus to show that the dip width is dependent on effective longitudinal correlations and not just on spectral correlations. The consideration of effective longitudinal coherence introduces a new degree of freedom to control biphoton indistinguishability without adding complexity to the system, thus offering potential benefits for quantum protocols. 
The LSC in this study is varied through two methods: adjusting the angular width of the pump beam and manipulating the longitudinal shift in the interfering pump beams before biphoton generation. 

\begin{figure*}
\centering\includegraphics[width=17.8cm,height=7.4cm]{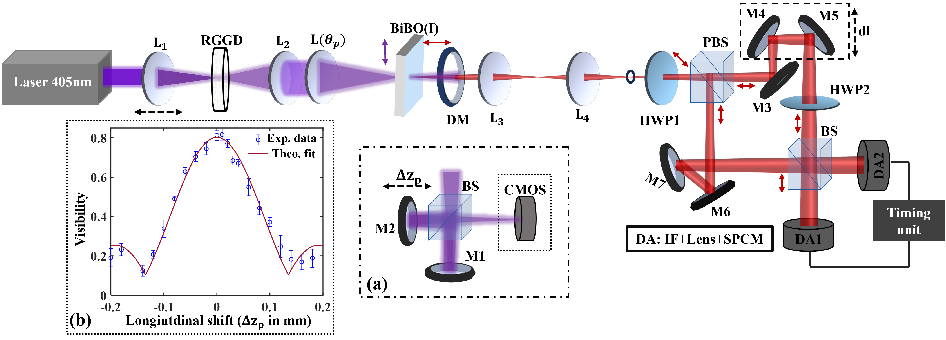}
\caption{HOM interferometric scheme to measure LSC in biphotons generated in collinear configuration with LSC pump beam (a) Michelson setup to vary $\Delta z_{p}$  after $\text{L}(\theta_{p})$ (b)Visibility of interference pattern of LSC pump beam w.r.t $\Delta z_{p}$ using CMOS. Notations: L lens, HWP half-wave plate, M mirror, PBS polarization beam splitter, IF interference filter, DM dichroic mirror, BS  beam splitter, CMOS complementary metal-oxide semiconductor, SPCM single photon counting module and arrow of beam colour denotes polarization.}
\label{expt}
\end{figure*}
 The assembly of a quasi-monochromatic laser (405 nm) beam along with RGGD and lenses ($\text{L}_{1},\text{L}_{2}$ and $\text{L}(\theta_{p})$) shown in Fig. 1, introduces the LSC in biphotons generated through BiBO crystal \cite{preeti2023}. 
 The Michelson interferometer setup in inset (a) is used as an additional tool to manipulate the LSC of biphotons by varying the interfering fields ($\Delta z_p$). The same setup verifies the LSC length ($\rho_{\parallel}$) through the visibility plot for the respective angular width ($\theta_p $= 0.06 rad), given in inset (b), and hence characterizes the pump parameters. 
 Further, dichroic mirror (DM) filtered out the collinearly propagating horizontally polarized biphotons emitted from type-I crystal.
 The orthogonally polarized biphotons coming from HWP and PBS traveled different path lengths (dl) in the HOM setup and efficiently collected using detection assemblies (DA1 and DA2).
 
The treatment of HOM interference between signal and idler with longitudinal spatially coherent pump is as follows: The state of collinearly propagating down-converted photons with identical polarization generated from the crystal is given by
\begingroup
\setlength{\abovedisplayskip}{5pt} 
\setlength{\abovedisplayshortskip}{5pt}
\setlength{\belowdisplayskip}{5pt}
\setlength{\belowdisplayshortskip}{5pt}
\begin{equation}
 |\psi \rangle =A_{0}\iint d\omega_{s}d\omega_{i} E_{p}(\omega_{p},z) \phi(\omega_{s},\omega_{i})|\omega_{s} \rangle |\omega_{i} \rangle
\end{equation}
\endgroup
where $A_{0}$ is a proportionality constant, $\omega_{s}, \omega_{i}$ and $\omega_{p}$ are angular frequencies of signal, idler, and pump, respectively, $\omega_{p}=\omega_{s}+\omega_{i}$, $E_{p}(\omega_{p},z)$ is the pump field, and $\phi=\sinc(\Delta k L/2)$ is phase matching function with $L$ as the crystal length and $\Delta k$ as phase mismatch term. Given the consideration of a quasi-monochromatic incoherent pump, we have selectively disregarded terms associated with the spectral width of the pump beam, concentrating solely on angular contributions. Since the pump is incoherent, the transverse spatial correlations would also be degraded owing to its dependence on the transverse coherence of the pump \cite{preeti2023}. 

A delay is introduced in the path of the signal photon such that it is incident on beam splitter $\tau=\frac{dl}{c}$ times later than the idler photon, as illustrated in Figure \ref{expt}. Consequently, the state of the generated photons after the HOM interferometer incorporates this delay between the signal and idler photons, along with the longitudinal shift in the pump. Thus, the effective longitudinal shift in this setup is given by $\Delta z = \Delta z_{p} + \text{dl}$, where $\Delta z_{p}$ represents the shift induced by the Michelson interferometer in the pump and $\text{dl}$ is the adjustable longitudinal shift introduced in the biphotons within the HOM interferometer.

After following beam splitter treatment, the resulting coincidence probability between the output of DA1 and DA2 is expressed as
\begingroup
\setlength{\abovedisplayskip}{5pt} 
\setlength{\abovedisplayshortskip}{5pt}
\setlength{\belowdisplayskip}{5pt}
\setlength{\belowdisplayshortskip}{5pt}
\begin{equation}
\begin{split}
 & g^{(2)}(t_{1},t_{2})\\
 &=  \frac{1}{2}\Big[g^{(2)}(t_{s},t_{i})- \iint d\omega_{s}d\omega_{i} \Gamma (\Delta z) \phi^{*}(\omega_{s},\omega_{i}) \phi(\omega_{s},\omega_{i}) \\
 &\times \exp \Big(- 2\big( \omega_{s}- \omega_{i} \big)\text{dl}/c \Big)\Big]. \label{actualP}
\end{split}
\end{equation}
\endgroup
where $ g^{(2)}(t_{s},t_{i})=  \iint d\omega_{s}d\omega_{i} \Gamma_{p} (\Delta z) \phi^{*}(\omega_{s},\omega_{i}) \phi(\omega_{s},\omega_{i})$, 
with 
\begingroup
\setlength{\abovedisplayskip}{5pt} 
\setlength{\abovedisplayshortskip}{5pt}
\setlength{\belowdisplayskip}{5pt}
\setlength{\belowdisplayshortskip}{5pt}
\begin{equation}
\begin{split}
 \Gamma_{p} (\Delta z)&=\langle E_{p}^{*}(\omega_{p},z)E_{p}(\omega_{p},z+\Delta z)\rangle\\ &= W_0 \int_0^{\theta_p} \exp(\iota k \Delta z (1 - \alpha^2/2) \alpha d\alpha\\
& = W_{0} \Delta k_{pz} \frac{\sin(\Delta k_{pz} \Delta z/2)}{\Delta k_{pz} \Delta z/2}  \exp (\iota k_{p0z}\Delta z)
\end{split}
\end{equation}
\endgroup
is the contribution from the pump envelope function. 
Here, $\alpha$ denotes an arbitrary half of the angular width of the pump beam, $k_{p0z}=\frac{\omega_{p}}{c} \cos^{2}\big(\frac{\theta_{p}}{2}\big)$, $\Delta k_{pz}=2\frac{\omega_{p}}{c}\sin^{2}\big(\frac{\theta_{p}}{2}\big)$ and $\rho_{||}=\lambda_{p}/\Big(2\sin^{2}\big(\frac{\theta_{p}}{2}\big)\Big)$. With approximation for small angles, the longitudinal spatial coherence is expressed as $\rho_{||}=2\lambda_{p}/\theta_{p}^{2}$. 
The incorporation of spectral filters along the signal and idler paths restricts the transmitted frequency range. Consequently, in the aforementioned equation, the spectral bandwidth is replaced by the constant bandwidth of the spectral filter used in the setup. This allows for isolating and deducing the effect of longitudinal spatial coherence in biphotons.
\begin{figure}[ht!]     \centering\includegraphics[width=.8\linewidth]{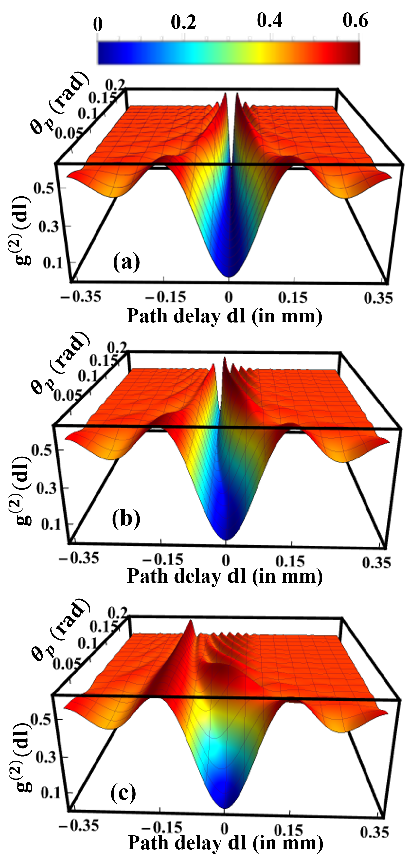}    \caption{  HOM interference depiction in 3D plot of $g^{(2)}\text{(dl)}$ with variation in path delay (dl) and $\theta_{p}$ for (a) $\Delta z_{p}=0$ mm (b) $\Delta z_{p}=0.05$ mm and (c) $\Delta z_{p}=0.15$ mm. 
 }\label{HOMTHEORY} \end{figure}
 \begin{figure*}\centering \includegraphics[width=17.8cm,height=10cm]{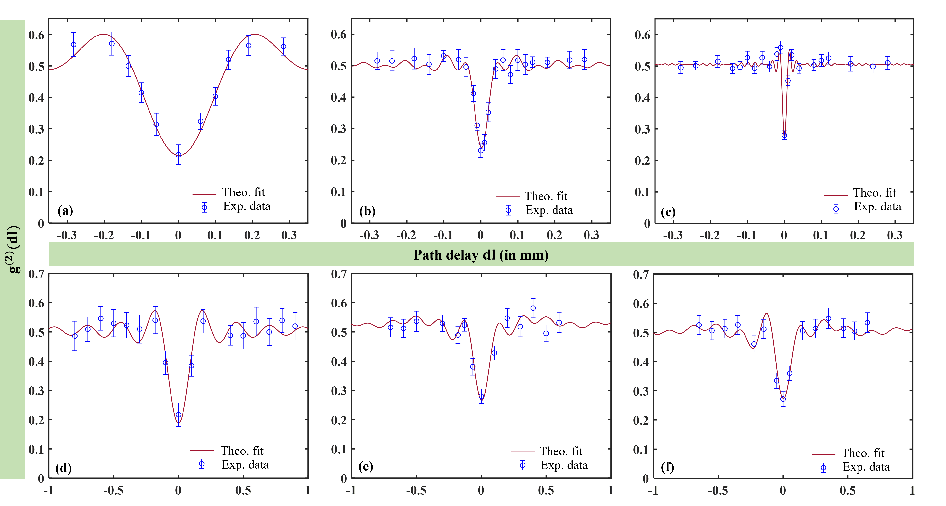} \caption{Plot of second-order correlation function with variation in path delay in signal: for pump with angular divergence (a) $\theta_p$ =0.06 rad (b) $\theta_p$ =0.117 rad  (c) $\theta_p$ =0.235 rad with $\Delta z_{p}$ =0 mm; and for longitudinal shift in pump (d) $\Delta z_p$ =0 mm (e) $\Delta z_p$ =0.05 mm (f) $\Delta z_p$ =0.15 mm with $\theta_{p}$=0.06 rad.} \label{Exptangular} \end{figure*}

 \begin{figure}[hb!]     \centering\includegraphics[width=.9\columnwidth]{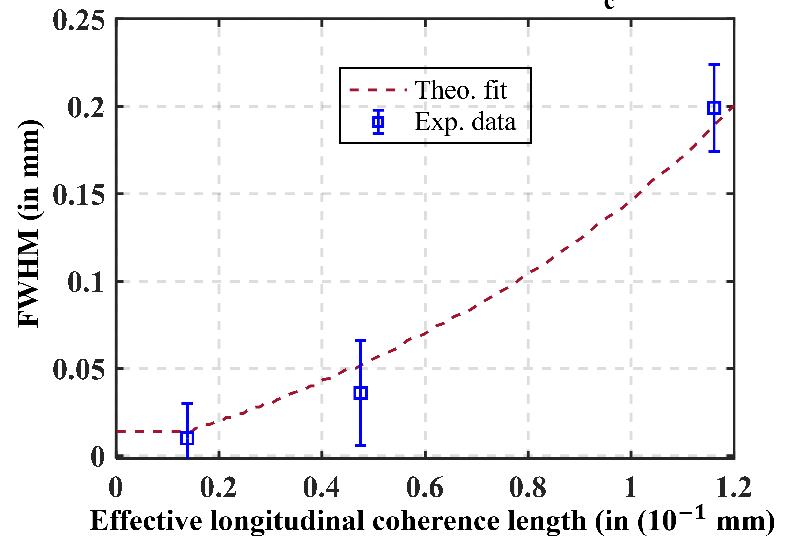}    \caption{Variation in FWHM of HOM interference dip with effective longitudinal coherence length ($L_{c}$) of pump.
 }\label{FWHM} \end{figure}
 
The theoretical analysis of HOM interference between signal and idler photons with variations in angular divergence and longitudinal shift is shown in Fig. \ref{HOMTHEORY}. The figure depicts 3D plots of the second-order correlation function $g^{(2)}(\text{dl})$ as a function of path delay and angular width of the pump, with the subfigures corresponding to different values of $\Delta z_{p}$. The width of the HOM dip is influenced by the effective longitudinal coherence, which is dependent on spectral bandwidth of the detection filter and the longitudinal spatial coherence of the interfering photons. The ripples around the central dip arise from both the spectral bandwidth of the interference filter and the pump envelope function. Since the spectral filter is constant across all cases, the dip is primarily affected by the longitudinal coherence length of the photons. The subfigure (a) shows that for $\Delta z_{p}=0$ mm, the visibility of the HOM dip attains unity for all values of $\theta_{p}$, however as $\theta_{p}$ increases the width of dip decreases, indicating a reduction in indistinguishability and longitudinal spatial coherence length of biphotons, mirroring the pump's longitudinal spatial coherence. However, in Fig. \ref{HOMTHEORY}(b) and (c), an increment in $\Delta z_{p}$ reveals a shift in the interference pattern, including peak shifts and visibility degradation at higher $\theta_{p}$ values. 
This decrease in visibility parallels fringe visibility behavior for LSC classical beams in a Michelson interferometer. The figure implies that photon indistinguishability can be altered without changing the spectral bandwidth of the pump.

The plots of $g^{(2)}$(dl) as a function of dl for different values of the pump angle $\theta_{p}$  are shown in the upper subfigures (a-c) of Fig. \ref{Exptangular}, corresponding to the experimental setup illustrated in Fig. \ref{expt}. 
Experimental data points, represented by circles, are accompanied by capped vertical lines indicating the standard deviation of the measurements. The solid lines depict theoretical curves fitting the experimental data.
The results indicate that dip width decreases with increasing angular divergence of the pump, attributed to a reduction in the longitudinal spatial coherence length of the biphotons while maintaining a fixed spectral bandwidth. This reduction parallels the decrease in the longitudinal spatial coherence (LSC) length $\rho_{||}$ of the pump beam, which is inversely proportional to angular divergence. The FWHM of the dip as a function of the pump's effective longitudinal coherence length is shown in Fig. \ref{FWHM}. This is obtained by calculating the FWHM of the interference pattern described in Eq. (\ref{actualP}) for various angular divergences of the pump and plotting it against $L_{c}$, where 
$L_{c}$ is given by 
$L_{c}=\frac{(2l_{c}\lambda_{p}/\theta_{p}^{2})}{(l_{c}+2\lambda_{p}/\theta_{p}^{2})}$. The plot reveals an increase in FWHM, indicating a corresponding enhancement in the indistinguishability of the photons.  Thus, it is evident that the effective longitudinal coherence of the pump (combining LSC and temporal coherence) results in the manipulation of the indistinguishability of the generated biphotons.
The bottom subfigures (d-f) of Fig. \ref{Exptangular} display $g^{(2)}$(dl) for varying $\Delta z_{p}$ values at a fixed pump angle of 
$\theta_{p}=0.06$ rad. Here, dip width decreases with increasing $\Delta z_{p}$, accompanied by reduced visibility. The lower experimental visibility compared to theoretical predictions may stem from imperfections in photon sources, single photon detectors in free-space configuration, and optical components. The observed behavior aligns with theoretical expectations: SPDC photons generated with a larger $\Delta z_{p}$ exhibit smaller dip widths. Notably, for 
$\Delta z_{p}=0$ mm, high visibility indicates indistinguishable photons, akin to fringe visibility in a Michelson interferometer (Fig. \ref{expt}(b)). This suggests that the longitudinal spatial coherence property of the pump is effectively transferred to the generated photons.

The manipulation of spatial and temporal coherence in pump fields has emerged as a critical method for tailoring the correlations of biphotons, particularly concerning their applications in quantum spectroscopy, microscopy, and imaging. This article introduces the use of LSC, which has already been proven effective in classical optical tomography and imaging, to address the resolution limitations of monochromatic beams in microscopy and OCT due to large coherence lengths. By employing LSC with a quasi-monochromatic pump, this work enables quantum microscopy and QOCT by tailoring photon indistinguishability. Additionally, encoding information in the angular width of the pump could allow for selective molecular excitation, advancing quantum sensing and imaging. Integrating LSC with homogeneous polarization further facilitates multiplex encoding for efficient parallel processing. The use of LSC with quasi-monochromatic pumps exhibits significant opportunities for innovations in quantum imaging, sensing, tomography, and computation. 

In conclusion, the study reveals that the wavepacket overlap of photons in the longitudinal direction is not entirely dependent on their spectral properties (frequency-width). Instead, it is significantly dependent on the effective longitudinal coherence of photons, which can be altered with either the change in temporal coherence or longitudinal spatial coherence. In this study, we restricted the variation in spectral bandwidth of photons by choosing quasi-monochromatic pump beam in the SPDC process. Despite this, the correlation width of photons, revealed from HOM interference dip still showed the variation with the change in the LSC features of the pump. The results demonstrate the dependence of two-photon interference on angular divergence and longitudinal shift of the pump beam. 
Additionally, the findings strongly suggest that the LSC property of the pump is effectively transferred to the generated photons implying the control in LSC of photons through control of angular spectrum of pump beam. 
The valuable insights of this study about the interplay between the longitudinal spatial coherence of the pump beam and the characteristics of the generated photons through SPDC process, thereby contributing to a fundamental understanding of quantum interference and laying important groundwork for future applications in quantum tomography and spectroscopy.
\begin{acknowledgments}
 The authors acknowledge the funding from the Department of Science and Technology (DST), India under the QUEST scheme (DST/ICPS/QUEST/Theme-I/2019) to carry out this work.  
\end{acknowledgments}
%
 
\end{document}